\documentclass[11pt]{amsart}

\usepackage[latin1]{inputenc}
\usepackage{amsmath,amssymb,amsfonts,amsthm,mathtools,dsfont}
\usepackage{color,verbatim,enumitem,psfrag,bbm,comment}
\usepackage{wrapfig}
\usepackage{graphicx}
\usepackage{caption}
\usepackage{subcaption}
\def\R{{\mathbb R}}

\begin{document}

\title{End-to-End Optimization of High Throughput DNA Sequencing}

\author{E. O'Reilly}
\address{Department of Mathematics, University of Texas, Austin}
\email{eoreilly@math.utexas.edu}
\author{F. Baccelli}
\address{Departments of Mathematics and of ECE,
University of Texas, Austin}
\email{baccelli@math.utexas.edu}
\author{G. de Veciana}
\address{Department of ECE, University of Texas, Austin}
\email{gustavo@ece.utexas.edu}
\author{H. Vikalo}
\address{Department of ECE, University of Texas, Austin}
\email{hvikalo@ece.utexas.edu}

\begin{abstract}
At the core of high throughput DNA sequencing platforms lies a bio-physical surface 
process that results in a random geometry of clusters of homogenous short DNA 
fragments typically hundreds of base pairs long.  -- {\em bridge amplification}.
The statistical properties of this random process and length of the fragments are critical as they affect 
the information that can be subsequently extracted, i.e., density of successfully inferred 
DNA fragment reads. The ensemble of overlapping DNA fragment reads are then used 
to computationally reconstruct the much longer target genome sequence, e.g, ranging
from hundreds of thousands to billions of base pairs. The success of the reconstruction 
in turn depends on having 
a sufficiently large ensemble of DNA fragments that are sufficiently long. In this paper 
using stochastic geometry we model and optimize the end-to-end process linking and 
partially controlling the statistics of the physical processes to the success of the 
computational step. This provides, for the first time, a framework capturing salient 
features of such sequencing platforms that can be used to study cost, performance or 
sensitivity of the sequencing process.
\end{abstract}


\maketitle

\section{Introduction}

Rapid and affordable detection of the order of nucleotides in DNA mo\-le\-cu\-les
has become an indispensable research tool in molecular biology. In this paper,
we consider the most prevalent sequencing technology that relies on reversible 
terminator chemistry \cite{bent08} and the ``shot gun sequencing" strategy 
\cite{mess81,vent96,vent98} to determine 
long DNA strands. Our goal is to develop a model and an associated mathematical 
framework that enable optimization of the end-to-end cost of DNA sequencing. 
For this, we draw upon stochastic geometry and queueing-theoretic tools to 
model and analyze salient characteristics of growing DNA clusters on the surface 
of a sequencing chip and optimize the process of sequence assembly from the 
short reads provided by the sequencing device. These developments provide a
systematic basis to study the 
tradeoffs and maximize the cost efficiency of the sequencing procedure.
 
\subsection{High-level Description of the Problem}

A target DNA strand to be sequenced can be viewed as a possibly long, e.g. $10^9$, 
sequence of $L$ letters. In shot gun sequencing a large number of copies of the target
are first randomly cut into {\em fragments}; the fragments are then sequenced, each 
providing a {\em read} of length $l$, where $l$ is much smaller than $L$ \cite{vent96,vent98}. 
The approach 
involves two key steps. In Step 1, one reads as many fragments as possible -- as
we elaborate later in this section, that can be parallelized and therefore performed 
very efficiently. In Step 2 one attempts to reconstruct the target sequence by leveraging 
the library of overlapping reads obtained in Step 1 -- this step is referred to as {\em assembly}.
For clarity let us consider the two steps independently, although one should keep in 
mind that they are intimately linked and will, in the sequel, be jointly optimized. 

To functionalize the surface of a DNA chip (referred to as a {\em flow cell}), DNA fragments
are first scattered across its surface whereupon they attach at random locations \cite{bent08}. A single
fragment is insufficient to generate a signal that is detectable; to remedy this, each of the
initially positioned fragments which we refer to as {\em germs} are replicated in parallel a 
number of times through the process called {\em bridge amplification.} The resulting ensembles of fragments, each comprising hundreds of identical copies of a germ, 
enable signal amplification and 
accurate DNA sequence detection. The germ replication can be viewed as a spatial branching 
process happening on the surface of the flow cell. The result of each such process, in the 
simplest case, is roughly a disc of the fragment's copies centered at the location where the original 
strand (germ) happened to attach to the flow cell. The radius of the disc can be controlled by 
the number of steps of the bridge amplification process. As we shall see, due to possible 
interaction amongst such growth patterns, the resulting shapes might be more complex than 
discs, so in the sequel we will more generally refer to them as {\em clusters}. 

As mentioned above, each fragment is replicated to generate a cluster of identical molecules which
enables signal amplification and thus facilitates sequence detection, i.e., the underlying letter 
reading mechanism. Reads are obtained in parallel, i.e., all clusters are read simultaneously 
one letter at a time. This is accomplished by relying on reversible terminator chemistry \cite{bent08} where the
first unread letter of each fragment is identified by detecting the color of the fluorescent 
label attached to the nucleotide bound to it.\footnote{More precisely, a mixture of free nucleotides each labeled with one out of four possible
fluorescent tags is added to the flow cell; a complementary nucleotide attaches to the 
topmost unpaired base (unread letter) in each DNA fragment of the flow cell. The free 
nucleotides are modified in such a way that once incorporated they terminate the 
sequence extension, i.e., a modified nucleotide incorporated into a strand prevents 
further strand extension. 
Once the fluorescent tag is identified, the nucleotide's modification (i.e., its terminating
property) is ``reversed" and incorporation may proceed further.} Since clusters contain 
multiple copies of the fragments, with proper illumination each cluster will light up the color 
indicative of the latest letter/base being examined. This process, referred to as {\em 
sequencing-by-synthesis}, is applied sequentially for say $l$ steps and thus in principle 
one can determine the first $l$ letters of all fragments on the flow cell -- this is the aim 
of Step 1.

There are two caveats however. First, the successive reading of nucleotides can fail 
for some fragments in each cluster. Specifically, on a given step, say $k$, the 
chemical processes associated with reading the $k$th nucleotide base may fail or 
jump ahead to the $(k+1)$st one.  Thus at each step only a fraction of the fragments 
in a cluster have their $k$th nucleotide properly marked with the correct fluorescent 
marker. As this proceeds, an increasing number of markers get {\em out of phase}, 
i.e., the disc/cluster will eventually appear to have a mix of colors, making correct 
detection of the fragment's next letter increasingly difficult. 
To deal with this phenomenon, typically
referred to as {\em phasing}, a number of {\em base calling} methods have been
proposed in recent years \cite{erli08,kirc09,kao09,kao09b,das12,das13}.
Indeed, amplifying the
signal is the reason for synthesizing the cluster of duplicates of each fragment in the
first place, i.e., clusters are meant to enable in-phase addition of light emanating 
from a number of identical fluorescent tags. 


The second caveat is that randomly placed germs may be grown into clusters that 
overlap which will also impair the reading process. Larger clusters will tend to 
experience more overlaps. In essence, this is a random disc packing problem:
if two germs happen to attach to the flow cell at distance $r$ from each 
other, any cluster growth that leads to discs of radius larger than $r/2$ leads to such
an overlap and hence to an impaired reading of the letters of the corresponding two 
strands.

We are now in a position to articulate the main tradeoffs that drive the efficiency of 
shotgun sequencing which assembles the target using short reads from a flow cell.
There are three main parameters at play: the density $\lambda$ of fragments
initially placed on the flow cell, the length $l$ of the fragment reads, 
and the disc radius $r$ associated with the bridge amplification process:
\begin{itemize}
\item $\lambda$ large looks desirable (because more fragment reads
will facilitate Step 2) but could be problematic for
any fixed $r$ because of possible disc overlaps; 
\item $l$ large is desirable (because it will facilitate Step 2)
but could be problematic because of the deterioration in reads quality
as base pair reads get out of phase; 
\item $r$ large is necessary (to amplify signals and facilitate detection) 
but this precludes the desire for a large $\lambda$ (because of disc overlaps).
\end{itemize}
We pose the following question: what is the optimal set of parameters
to maximize the ``yield'' and/or possibly minimize the cost 
in the presence of all these tradeoffs?

There are many ways to frame the problem of optimizing yield in such systems.
In this paper we will first consider optimizing fragment yield, which 
we define as the density of length $l$ fragments successfully read 
per unit area of the flow cell. Then we consider 
metrics that are more directly tied to the final objective 
of sequencing length $L$ target DNA sequence. To that end let us 
consider Step 2, the reassembly process. 

Following a simple, stylized model of the process, a condition for successful 
reassembly is that the collection of successfully read fragments {\em covers} 
the target DNA -- i.e., if one denotes by $a_1,\cdots,a_L$ the target DNA sequence, 
then in the set of correctly read fragments there should exist
a subset of fragments, say $s_1,\ldots,s_k$ such that 
$s_1$ contains the sequence $a_1,\ldots a_{p_1}$, 
$s_2$ contains the sequence $a_{p_1+1},\ldots a_{p_2},\ldots$, 
and so on, until $s_k$ contains the sequence $a_{p_{k-1}+1},\ldots a_{L},$
for some $p_1,\ldots,p_{k-1}$. This stylized model for reassembly is highly 
simplified. In practice we may encounter two scenarios, 
{\em de novo} and {\em reference-based} sequencing.
De novo sequencing \cite{myer95, idur95,mill10} refers to sequencing genomes which have not
been previously characterized, while reference-based
sequencing \cite{delc02} refers to the sequencing task where one or more previously
sequenced references which are similar to the target are available.
These present different challenges in the reassembly process, i.e., 
determining where fragments should be placed to reconstruct the target
(sequence alignment has attracted considerable amount of attention,
e.g., see \cite{lang09,durb09,durb10,li08maq,lunt11,ruff11} and the
references therein).
Moreover, reassembly requires additional slackness in the cover,
i.e., in $s_i$, there should be enough letters on the left
of $a_{p_{i-1}+1}$ and on the right of $a_{p_{i}}$ to
correctly reconstruct the long sequence from the fragments
exactly as in a puzzle. Assuming the ability to appropriately map/align 
fragments, the existence of a covering is a minimal
requirement to sequence the target, see \cite{BBT13,MBT13} and
references therein for modern discussion of this problem.

A critical tradeoff associated with Step 2 now emerges. It is between the number 
of correctly read fragments and their length. A large collection of fragments is 
helpful, but if they are too short, i.e., $l$ is small, it is difficult to obtain a covering 
of the entire DNA target sequence\footnote{Additionally, in the reference-guided 
assembly scenario, longer fragments are easier to align with the target.}.

This brings us to the main problem addressed (and solved) in this paper.
We set as our goal the determination of the parameters associated with
the sequencing process as described earlier
(namely the density $\lambda$ of fragments placed on the flow cell,
the duration of the growth process, and the length of the fragment reads $l$) 
which ensure a pre-specified probability,
say $\delta$, e.g, $99\%$, of obtaining a covering of the target DNA sequence 
at {\em minimal operation cost}.
Operational costs can be organized in two main categories:
(a) those associated with raw materials, e.g., 
DNA copies, reagents, flow cells; (b) those associated with time,
e.g., the time spend on the sequencing machine or
the execution time of the signal processing algorithms.  
In this paper we shall for simplicity
adopt the {\em flow cell area} as the cost. 
Some reagent costs and some flow cell processing time costs 
might be proportional\footnote{In this discussion, when we say
proportional, we mean proportional through multiplicative factors
that do {\em not depend on the optimization variables}, namely on
the length of the strands, the density of the strands on the flow cell,
or the duplication factor.} to the flow cell area.
We find it important to stress that our
general mathematical framework can also accommodate other costs.
For instance costs that are proportional to the total number of
strands (this might be the case for certain processing times)
or to the number of DNA copies (raw material), rather than to the area.
We will not discuss them here for the sake of brevity.

\subsection{Related Work}


To our knowledge, this is the first work attempting to model and analyze the cluster 
growth process with a view on optimizing DNA sequencing cost/yield. The detailed 
simulations of the surface physics associated with the bridge amplification process, 
\cite{MSM03,MES05}, support that the disc/cluster processes we introduced earlier 
and will use are well suited. Work optimizing this process has taken place in industry 
where empirical evidence and simple rules of thumb have been used.
There is, however, a substantial body of work towards 
developing mathematical tools for analyzing random spatial processes (see, e.g., 
\cite{Stoyan} and some of our work \cite{Baccellig}). 
Indeed, this branch of mathematics is now ubiquitous with applications in
material science, cosmology, life sciences, information theory, to name a few.
Further
developments of the mathematical foundations have recently been carried out by us 
in \cite{Baccellig}, and proven to be invaluable, e.g., to understand fundamental 
characteristics of large wireless systems and optimizing their performance. Such 
stochastic geometry models have been embraced by academia and industry, 
providing insight into current and future technological developments. Indeed the aim 
of this paper is to show that this may also play a role in the DNA sequencing setting. 

As mentioned earlier in this section, sequence assembly may be performed with or 
without referring to a previously determined sequence (genome, transcriptome). 
{\it De novo} genome shotgun assembly is a computationally challenging task due 
to the presence of perfect repeat regions in the target and by limited 
lengths and accuracy of the reads \cite{mill10,BBT13,MBT13}. In the {\em reference-guided} assembly 
setting, the reads are first aligned (i.e., mapped) to the reference, easing some of 
the difficulties faced by the {\em de novo} assembly \cite{delc02}. However, the reference 
often contains errors and gaps, creating a different set of challenges and problems. 
In fact, if the sample is highly divergent from the reference or if the reference is missing 
large regions, it may even be preferable to use {\em de novo} assembly \cite{iqba12}.
While the development of methods for sequence assembly received significant
attention, ultimate limits of their performance have been less explored. The pioneering 
work of \cite{LaW88} provided the first simple mathematical model for the reassembly 
process. This has been followed with various refinements \cite{BBT13,MBT13} but to 
our knowledge none has provided a mathematical framework to compute the flow cell 
parameters needed to achieve a given likelihood of full target coverage. Moreover, no 
previous work has linked this end objective to the optimization of the end-to-end process 
as we will do in this paper. 


\subsection{Organization of the paper.}
In Section \ref{sec:mod} we propose a stochastic geometric model for the distribution 
of clusters on the flow cell resulting from the bridge amplification process. The 
proposed model enables analysis of the impact of the geometry of clusters
on the achievable yield of fragment reads.  Fragment read yield optimization is 
considered in Section \ref{sec:yield}. In Section \ref{sec:e2e} we propose a simplified 
model for the reassembly problem which is related to a queueing model and analysis. 
This allows us to consider the end-to-end
cost optimization of the sequencing process to meet a desired 
likelihood of coverage for the target DNA sequence which is carried out  
in Section \ref{sec:e2e}.

\section{Stochastic geometric models for shotgun DNA sequencing}
\label{sec:mod}

In this section, we introduce basic geometric models for the cluster
processes associated with the DNA fragments resulting from 
bridge amplification procedure on the surface of the flow cell. 
These are the {\em singleton cluster}, {\em shot-noise}
and the {\em Voronoi } models, respectively.
These processes will be tied to the salient features of 
fragment reading mechanisms. 

In the singleton cluster process model, all clusters that
intersect (or touch) another cluster are discarded. 
The retained clusters are roughly modeled as
discs of radius $r$ consisting of duplicates of the same DNA fragment. 
In the shot-noise process model an attempt is made to read 
each cluster. Isolated clusters are as in the singleton cluster case. A cluster
which is in contact with one or more clusters is still
analyzed as a disc of radius $r$; however, depending on the
number and shape of the other clusters in contact, part
of the light signal stemming from that disc
creates an interference which is treated as noise.
If signal dominates interference/noise, one can still
read this cluster.
Finally, the Voronoi case studies the (hypothetical and
somewhat futuristic) scenario where one computes optimal
masks that allow one to mask all clusters that are in contact
with the tagged cluster and hence to cancel interference.

Some of these models will be used
in Section \ref{sec:e2e} for the reassembly
optimization alluded to above. 
For each case, we describe the mathematical approach
used to evaluate its performance. This
will also be used in order to
give some yield optimizations of independent interest
in Section \ref{sec:yield}.

\subsection{Random seed model and growth model}

We consider the locations of the initial DNA fragments
(the centers of the clusters or the seeds) to be a homogenous
Poisson point process $N$ on $\R^2$, with intensity $\lambda$. This
parameter is simply the number of seeds per unit area.
Growth of clusters is assumed to be radially homogenous, so if a 
cluster does not come into contact with any other cluster over a time $r$
of growth, it will form a disc of radius $r$. 
The amplification process creates up to 1000 copies of the initial
fragment in a disc of radius $.5$ microns \cite{IlluminaWhitePaper}.
This gives us a density $a$ of $\frac{4000}{\pi}$ fragments per
square micron.

If it does come into
contact with another cluster, we assume the growth in that direction
stops, but growth continues in all other directions. As $r$
approaches infinity, the configuration of clusters becomes
the {\em Voronoi diagram} associated with the point process $N$ \cite{Stoyan}.
For $r$ finite, the shapes are known as the {\em Johnson-Mehl growth model}
\cite{Stoyan}.

\subsection{Read reliability model}

As already explained,
phasing problems occur because nucleotides occasionally fail to
incorporate in particular duplicates or anneal to the base pair
right next in line. These duplicates are then out of sync with the 
rest of the duplicates and give off a different color signal.
So even though amplifying the fragments gives a much stronger signal,
there is noise due to these out-of-phase duplicates limiting the
accuracy of the reads.

To model this in a single cluster, let $X_l$ denote the number
of copies of the original DNA fragment that remain in phase after
$l$ steps of the process. Let $p$ be the probability that a DNA
fragment gets out of phase at one step. 

The random variable $X_l$ has a binomial distribution,
where the number of trials is the number of DNA fragment copies
in a cluster, and $(1-p)^l$ is the probability that a fragment
remains in phase, after $l$ steps, i.e. the probability
of success.

At first glance, it makes sense to require that
$p(l) = (1-p)^l > \frac{1}{2}$
(so that on average, more than half the duplicates remain in phase)
to have a correct read.
However, this does not capture certain phenomena, e.g. the fact that as the radius becomes
very small, fluctuations are high around the mean. 
We will hence require that the number of duplicates remaining
in phase is above half by some positive margin.
The probability of a correct read will
then rather be $\mathds{P}(X_l \geq \frac{a\pi r^2}{2} + \epsilon) $
for some positive epsilon, which is an important complementary
parameter of our model.
The value of epsilon used here is 10, since the output yields using this value are on the same order as the density of clusters achieved by Illumina technology \cite{IlluminaWhitePaper}.

\subsection{The singleton cluster model}

For a cluster with unimpeded growth over time $r$,
the number of DNA fragments in a cluster is $a\pi r^2$, so
$X_l \sim B(a\pi r^2, p(l))$. 

When the number of trials is large, the binomial distribution
can be approximated by the normal distribution. In the singleton case:
\begin{center}
$X_l \sim \mathcal{N}(a\pi r^2p(l), \sqrt{a\pi r^2p(l) (1 - p(l))})$.
\end{center}

A cluster centered at $y$ in $N$ is isolated after time $r$ if the ball centered at $y$ with radius $2r$ does not contain any other point of N. A homogenous Poisson point process is stationary, so we can consider a typical ball centered at 0. 
Given intensity $\lambda$ and radius $r$, we can then calculate the intensity of isolated clusters $\lambda_i(\lambda, r)$. By Slivnyak's theorem,
\begin{align*}
\lambda_i(\lambda, r) &= \lambda \mathds{P}^0( N^! (B(0,2r)) = 0) \\
&= \lambda \mathds{P}( N(B(0,2r)) = 0) \\
&= \lambda e^{-4\lambda\pi r^2}.
\end{align*}

The overall fragment yield of singletons is the intensity of
the isolated clusters times the probability a cluster will  
enjoy a correct read:
\begin{center}
 $\lambda_y(\lambda, r, l) = \lambda e^{-4\lambda\pi r^2} \mathds{P}(X_l \geq \frac{a\pi r^2}{2} + \epsilon) $.
\end{center}

\subsection{The shot-noise model}
\begin{wrapfigure}{r}{0.44\textwidth}
\vspace{-20pt}
\hspace{-1cm} \includegraphics[scale=.20]{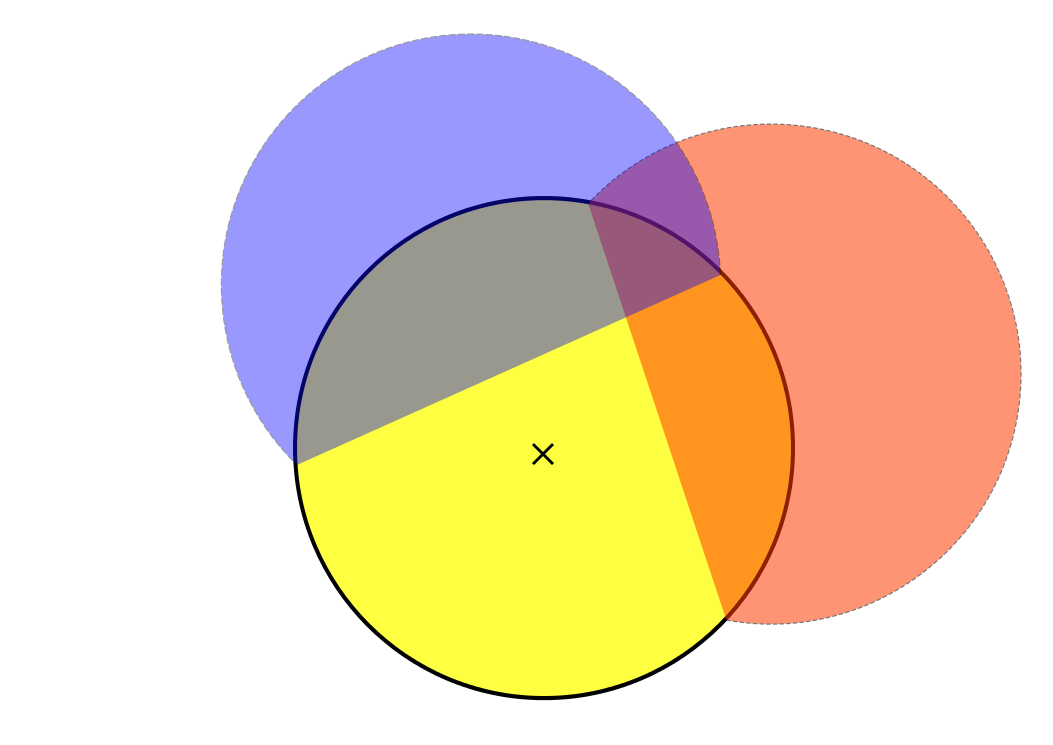}
\vspace{-20pt}
\caption[Packing]{Cluster interference.}
\vspace{-10pt}
\label{fig1}
\end{wrapfigure}

Using only isolated clusters is clearly suboptimal.
We consider here the situation where all clusters are used.
In this case, for each cluster, the amount of interference
from contact with other clusters during the growth/duplication
process has to be taken into account.
According to our growth assumptions, the area of the typical cluster
(assumed with a seed located at 0) is
$|V_0 \cap B(0,r)|$, where $| \cdot |$ is Lebesgue measure, $V_0$ is
the Voronoi cell of the point at 0, and $B(x,r)$ the ball
of center $x$ and radius $r$.
Here, we use a lower bound for the area that is easier to calculate.
The interference encountered from another cluster centered
at $x_i \in N$ is considered to be half of the area of the
overlap between the discs $B(0,r)$ and $B(x_i, r)$. 
We take the total interference for the typical cluster to be the sum
of these areas over all surrounding clusters. This is in fact an upper bound on the actual interference, e.g. triple intersections are counted twice (see Figure \ref{fig1}). 

This interference upper bound can be described in terms of a shot-noise field $I_N$ defined on $\mathds{R}^2$ as a functional of our Poisson point process $N$ with response function $\alpha_r(x) = \frac{1}{2}|B(0,r) \cap B(x, r)|$. For fixed $r$, $a_r$ depends only on the distance $||x||$, so we write $a_r(||x||)$.
The total interference is $I \equiv I_N = \int_{\mathds{R}^2} a_r(||x||) N(dx)$.

The Laplace functional of the interference is:
\begin{center}
$L_{I}(s) = e^{-\int_{\mathds{R}^2} (1 - e^{-sa_r(||x||)}) \lambda(dx)}$.
\end{center}
Overlap with the typical cluster $B(0, r)$ only occurs for clusters with centers contained in the ball of radius $2r$ around 0, so we consider only the interference on $B(0, 2r)$. Switching to polar coordinates, the Laplace transform becomes
\begin{align*}
L_{I}(s) &= \exp(-2\pi\lambda \int_{0}^{2r} (1-e^{-s\alpha_r(v)}) v dv),
\end{align*}
where
\begin{center}
$\alpha_r(v) = r^2\cos^{-1}\left(\frac{v}{2r}\right) - \frac{v}{2}\sqrt{r^2 - \frac{v^2}{4}}$.
\end{center}
The random variable $X_l$ (the number
of copies of the original DNA fragment that remain in phase after
$l$ steps)
is now binomial with the number of trials depending on the interference $I$.
Since the interference comes from fragments that are not copies
of the seed of the tagged cluster, this area contains no potential
in-phase fragments. The area occupied by fragments of interest
is $\pi r^2 - I$ and thus,
\begin{center}
$X_l(I) \sim Bin(a(\pi r^2 - I), p(l))$, where $p(l) = (1-p)^l$.
\end{center}
The cluster is still analyzed as a disk of radius $r$, so
the number of in-phase DNA fragments needed for a correct read
remains at least $\frac{1}{2}a\pi r^2 + \epsilon$.
The probability of a correct reading given interference $I = x$ is
$$g(x) = P(X_l(x) > \frac{1}{2}a\pi r^2 + \epsilon | I = x)$$ 
and the fragment yield is 
$$ \lambda_y(r, \lambda, l)= \lambda \int g(x) f(dx),$$
where $f$ is the law of $I$, which is known through its Laplace transform
\cite{Stoyan}.
The computation of this yield, which is based
on Fourier techniques, is discussed in Appendix \ref{a1}.

%
%
%
%
%

\subsection{The Voronoi model}
In this subsection, we consider an optimal scenario for collecting signal
reads using our assumptions about cluster growth. Clusters are allowed
to grow until they have formed a Voronoi tessellation. Then optimal
sized masks having the shape of each Voronoi cell are used to read the signal.
In this scenario no interference from neighboring clusters is present,
and the only variable to optimize is the intensity $\lambda$ of the
underlying point process.

The closed form of the distribution for the area of Voronoi cells is unknown, but it can be approximated by the generalized Gamma density:
\begin{center}
$f_{\lambda A}(x) = \frac{\gamma \chi^{\nu / \gamma}}{\Gamma(\nu / \gamma)}x^{\nu -1}\exp(-\chi x^{\nu})$ for $x \geq 0$.
\end{center}
This is the approximate distribution for the normalized cell size $\lambda A$, where $A$ denotes the 
area. For $\lambda = 1$, good choices for the area are: $\gamma = 1.08$, $\nu = 3.31$, and $\chi = 3.03$ \cite{Stoyan}.

For a general intensity $\lambda$ the area distribution is given by
\begin{center}
$f_{A}(x) = \lambda f_{\lambda A} ( \lambda x) = \lambda \frac{\gamma \chi^{\nu / \gamma}}{\Gamma(\nu / \gamma)}(\lambda x)^{\nu -1}\exp(-\chi \left(\lambda x\right)^{\nu})$ for $x \geq 0$,
\end{center}
where $\gamma = 1.08$, $\nu = 3.31$, and $\chi = 3.03$.

The expected fragment yield for the Voronoi case is then
\begin{center}
$\lambda_y(\lambda, l) = \lambda \int_0^{\infty} \mathds{P}(X_l > \frac{a A}{2} + \epsilon | A = x) f_A(x) dx$.
\end{center}

\section{Fragment and letter yield }
\label{sec:yield}

In this section, we first study the {\em fragment yield},
namely the mean number of correctly read fragments per unit
space of the flow cell. We then study the {\em letter yield},
which is based on the number of letters correctly read.

\subsection{Numerical results on fragment yield}

Below, we use the mathematical expressions obtained above to
optimize the yield in all three models for $l$ fixed.
For the singleton cluster and the shot-noise model, 
we optimize over the radius $r$ and the intensity $\lambda$.
For the Voronoi model, the optimization is over $\lambda$.

Table 1 and Table 2 show optimal parameters and fragment yields for $l = 200$ and $l = 150$.

\begin{table}[ht]
\caption{Optimal Parameters for $l=200$}
\centering
\begin{tabular}{c c c c}
\hline 
Value & Singletons & Shot-noise & Voronoi  \\ [0.5ex]
\hline 
$\lambda$ & 1.3981 & 1.9143 & 3.7478 \\
r & .2386 & .2447 &   \\
F-Yield (per square micron) & .2844 & .4145 & 2.8422 \\
[1ex]
\hline
\end{tabular}
\end{table}


%

For $l=200$, a 45.7\% increase in optimal fragment yield 
can be obtained when 
considering all clusters and not just the singleton ones. This increase
in yield comes with a slighter larger radius and higher intensity
than the optimal parameters for singleton clusters.
This corresponds to increasing the amount of time for replication
and increasing the number of initial DNA fragments spread over the flow cell.
This will mean more clusters overlap with their neighbors, but many
more correct reads can still be made for clusters that only run into
their neighbors late in the growth stage.

\begin{table}[here]
\caption{Optimal Parameters for $l=150$}
\centering
\begin{tabular}{c c c c }
\hline 
Value & Singletons & Shot-noise & Voronoi   \\ [0.5ex]
\hline 
$\lambda$ & 3.2625  & 4.6031 & 8.6723  \\
$r$ &   .1562  &  .1686  &   \\
F-Yield (per micron$^2)$ &  .9146  &  1.5704    & 8.6108  \\
[1ex]
\hline
\end{tabular}
\end{table}

For each case, decreasing $l$ to 150 from 200 resulted in a smaller
optimal radius and a larger optimal intensity. With a smaller $l$,
a fragment is more likely to remain in-phase, making it is easier for
in-phase fragments to comprise half the cluster plus the fixed 
margin $\epsilon$. This allows clusters to be smaller and more densely packed to obtain a higher yield.

The percent increases in the fragment yield obtained
when switching to $l= 150$ are
\begin{itemize}
\item Singletons:  $221.59 \% $
\item Interference: $278.87 \%$
\item Voronoi: $202.96 \%$.
\end{itemize}

\subsection{Numerical results on letter yield}
In view of the differences between $l=200$ and $l=150$,
it makes sense to consider the optimal letter yield, namely
$l\lambda_y(l,r, \lambda)$, 
the mean number of letters correctly deciphered
per unit space. So, below, optimization takes place w.r.t. $l$ as well.

\begin{table}[here]
\caption{Optimal Parameters for Letter Yield}
\centering
\begin{tabular}{ c c c c}
\hline 
Value & Singletons & Shot-noise & Voronoi  \\ [0.5ex]
\hline 
$\lambda$ &  6.2320 & 9.5769   &  12.1160    \\
$r$  &  .1130  & .1233  &    \\
$l$ &  91.4570  & 86.1716  & 119.8302  \\
L-Yield (per micron$^2)$ & 179.8272  &  344.2525 & 1535.7  \\
[1ex]
\hline
\end{tabular}
\end{table}
We see that the optimizing value of $l$ is actually somewhere
around $100$, which is shorter than the read length
provided by, e.g., Illumina's HiSeq sequencing platforms.

The percent increases in the fragment yield obtained
when switching from $l= 150$ to the optimal $l$ are
\begin{itemize}
\item Singletons:  $31.07 \% $
\item Interference: $49.14 \%$
\item Voronoi: $18.89 \%$.
\end{itemize}
In this optimized setting, the letter yield of the shot-noise case
is close to twice that of the singleton cluster case.

\section{Reassembly model and optimization}
\label{sec:e2e}

This section is focused on the optimization of
the probability of reassembly of the original DNA sequence.

\subsection{Reassembly model}
The reassembly question can be formulated in terms of
\begin{itemize}
\item $n$ the number of fragments in the genomic library
(fragments correctly deciphered);  
\item $L$ the DNA sequence length in base pairs (letters);
for the human genome, $L = 3$ billion;
\item $l$ the length of the fragments in the same unit.
\end{itemize}
We see $L$ as a segment on the real line
and we assume that the fragment
starting points form a Poisson point
process on the real line with parameter $\lambda = \frac{n}{L}$.

As already explained
we are interested in 
the probability of complete reassembly. We will first reduce this to
the probability that {\em all} letters are covered.

\subsection{Analytic expression for the reassembly probability}
 
Within the Poisson setting described above,
this can be reduced to a queuing theory question:
consider an $M/D/\infty$ queue, namely a queue with Poisson
arrivals, an infinite number of servers and a constant service time
of length $l$. The probability of reassembly is then just the
probability that the busy period in such a queue exceeds $L$. The distribution of the busy period can be determined through its Laplace transform.

Let $N$ be a Poisson point process on $\R$ with parameter $\lambda$. 
Let $\{T_k\}_{k \in \mathds{N}}$ be the following random sequence of times:
\begin{align*}
T_0 &= 0, \\
T_1 &= \begin{cases}
l, & \text {if } N(0,l) = 0 \\
\max\limits_{x \in N,\text{ } x \leq l} x, & \text{ if } N(0,l) > 0
\end{cases}, \text{ and } \\
T_k &= \begin{cases}
T_{k-1} + l, & \text {if } N(T_{k-1},T_{k-1} + l) = 0 \\
\max\limits_{x \in N,\text{ } x \leq T_{k-1} + l} x_i, & \text{ if } N(T_{k-1},T_{k-1} + l)> 0
\end{cases}
, \text{ for } k > 1.
\end{align*}
The busy period of the queue is $T \equiv T_K$, defined by
\begin{center}
$T_K = \begin{cases}
 l, & \text {if } K = 1 \\
l + \sum_{i=1}^{K-1} (\xi_i | \xi_i < l), & \text{ if } K > 1
\end{cases}$,
\end{center}
where $K$ is the random variable $K = \min\{n \in \mathds{N} : T_n = l\}$ and the $\xi_i$ are i.i.d random variables with distribution equal to that of the last Poisson arrival in the interval $(0, l)$. 
We claim that the $\xi_i$ are equal in distribution to $l - (\eta | \eta < l)$, where $\eta \sim \exp(\lambda)$.
Indeed, the distribution of the distance of the first point
in a Poisson point process on $\R$ from 0 is the same looking
forward and looking backward.
By this and translation invariance,
\begin{align*}
l- \min\limits_{x \in N : x \in (0,l) } |x| &\stackrel{(d)}{=} l - \min\limits_{x \in N : x \in (-l,0) } |x_i| \\
&\stackrel{(d)}{=} \max \limits_{x \in N : x \in (-l,0) } x  +  l \\
&\stackrel{(d)}{=} \max \limits_{x \in N : x \in (l,0) } x.  
\end{align*}
Now, let $\eta \stackrel{(d)}{=} \sim \exp(\lambda) $. The Laplace Transform of $(\eta| \eta < l)$ is
\begin{align*}
\Psi_{\eta | \eta < l}(s) &= \mathds{E}(e^{-s\eta} | \eta < l) \\
&=\frac{ \mathds{E}(e^{-s\eta}1_{\{ \eta < l\}}) }{\mathds{P}(\eta < l)} \\
&=  \frac{1}{1 - e^{-\lambda l}}  \int_0^l \lambda e^{-(s+\lambda) x} dx \\
&=  \frac{1}{1 - e^{-\lambda l}} \left[\frac{\lambda}{s + \lambda} ( 1 -  e^{-(s + \lambda)l} )\right].
\end{align*}
Now, if $K = 1$, then $T_K = l$. So, $\mathds{E}[e^{-s T_K} | K = 1] = e^{-\lambda l}$. For $k > 1$,
\begin{align*}
\mathds{E}[e^{-s T_K} | K = k] &= \mathds{E}[e^{-s( l + \sum_{i=1}^{k-1}\left( l - (\eta_i | \eta_i < l)\right) )}] \\
&=  e^{-sl} \prod_{i=1}^{k-1} e^{-sl} \mathds{E}[ e^{s\eta_i} | \eta_i < l] \\
&= e^{-sl} \left( e^{-sl} \Psi_{\eta | \eta < l}(-s) \right)^{k-1} \\
&= e^{-sl} \left( \frac{\lambda}{\lambda - s} \left[\frac{e^{-sl}  -  e^{ - \lambda l}}{1 - e^{-\lambda l}} \right] \right)^{k-1}. 
\end{align*}
Since $K$ is a geometric random variable with success probability $p = \mathds{P}(N(0,l) = 0) = e^{-\lambda l}$, the Laplace transform of the busy period $T$ is
\begin{align*}
\Psi_{T}(s) &= \mathds{E}[e^{-s T}] \\
&= \sum _{k=1}^{\infty} \mathds{E}[e^{-s T_K} | K = k ]\mathds{P}(K = k) \\
&= \sum _{k=1}^{\infty} e^{-sl} \left( \frac{\lambda}{\lambda - s} \left[\frac{e^{-sl}  -  e^{ - \lambda l}}{1 - e^{-\lambda l}} \right] \right)^{k-1}   p(1-p)^{k-1} \\
&= e^{-\lambda l}e^{-sl} \sum _{k=1}^{\infty} \left[ \frac{\lambda}{\lambda - s} \left[ e^{-sl}  -  e^{ - \lambda l} \right] \right]^{k-1} \\
&= e^{-\lambda l}e^{-sl} \left[ \frac{1}{ 1 - \left(  \frac{\lambda}{\lambda - s} ( e^{-sl} - e^{-\lambda l}) \right)} \right] \\
&=  \frac{e^{-(\lambda +s) l}(\lambda - s)}{ \lambda - s -  \lambda( e^{-sl} - e^{-\lambda l})}.
\end{align*}
We want to compute $\mathds{P}(T > L)$, so we
invert the Laplace transform of the CDF,
which is easily calculated from the Laplace transform of the density:
\begin{align*}
\mathcal{L}_{F_T}(s) = \frac{1}{s}\Psi_T(s).
\end{align*}
The inversion is done numerically using the Euler Inversion method  \cite{Abate}.

\subsection{Reassembly optimization}

In our optimization, we ask the following question: For a given genome length, what is the smallest area the sequencing flow cell needs to have in order 
to get a high probability that the entire genome can be reassembled?

To answer this, let $A$ be the area (in square microns) of the flow cell where fragments are replicated and sequenced. Then, $n = A\lambda_l$, where $\lambda_l$ is the optimal yield per square micron for fragments of length $l$ as derived in Section \ref{sec:mod}. Assume $L$ is given. 
Then consider the $M/D/\infty$ queue with arrival rate $\frac{A\lambda_l}{L}$ and service time $l$.

For each $l$, we find the minimum
$A$ such that the probability of reassembly, $\mathds{P}(T > L)$, is greater
than some threshold. Finally, we optimize over $l$ to find the smallest required area.  
The optimal $l$ is the fragment length that requires the least flow cell area to obtain the desired probability of reassembly.

\begin{figure}
\centering
\caption{}
\begin{subfigure}[b]{0.65\textwidth}
\includegraphics[width = \textwidth ]{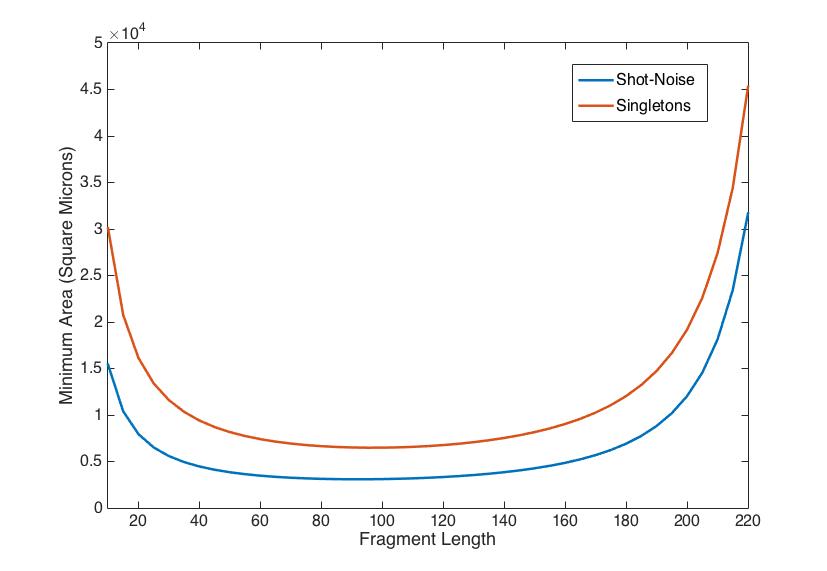}
\end{subfigure}
\begin{subfigure}{0.65\textwidth}
\includegraphics[width = \textwidth ]{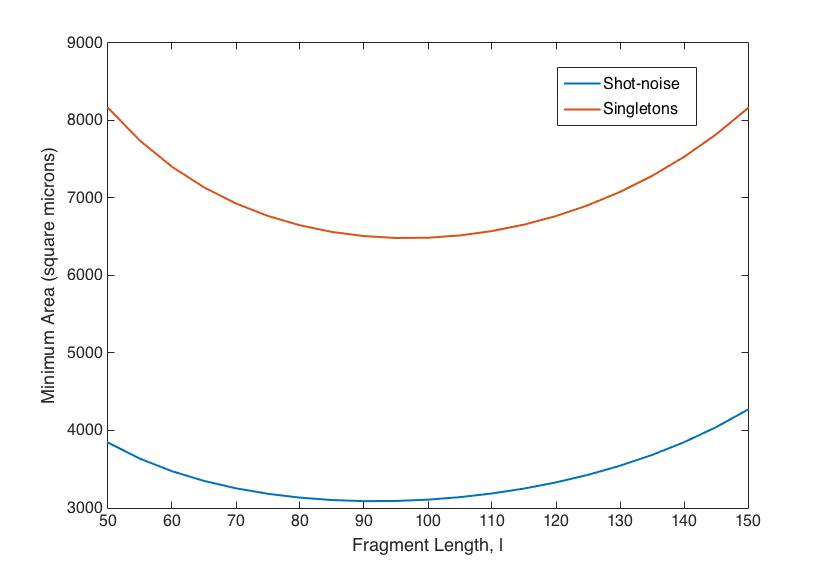}
\end{subfigure}
 \label{MinA}
\end{figure}

Figure \ref{MinA} shows the minimum area needed to achieve a probability of reassembly of .99
versus the length of the fragments for a genome of length $100,000$.

\begin{table}
\caption{Optimal Parameters needed for $\mathds{P}( T >  100,000) \geq .99$.}
\centering
\begin{tabular}{c c c}
\hline 
& Singletons & Shot-Noise \\ [0.5ex] \hline
Optimal $l$ & 97 & 92 \\
Minimum $A$ (square microns) & 6,483 & 3,088 \\
[1ex]
\hline
\end{tabular}
\end{table}

\section{Conclusion}
This paper establishes a connection,
which is new to the best of our knowledge,
between stochastic geometry and queuing theory on one side,
and fast DNA sequencing on the other side.
This connection allowed us to propose a simple model which 
captures the essential steps of the fast sequencing process:
segmentation of multiple copies of the DNA into random fragments,
replication of the randomly placed fragments on the flow cell,
spatial interactions between the resulting fragment clusters,
read of fragments through their cluster amplification, taking
the possibility of read errors and interference between clusters
into account, and finally assembly of the successfully read fragments.
This model is analytically tractable, which allowed us
to quantify and optimize various notions of yield, including
the yield of the end to end sequencing process, in function
of the key parameters.
This basic model seems generic and flexible enough for us 
to envision a series of increasingly
realistic and yet tractable variants for each step of the process
and eventually a comprehensive quantitative theory for this class of
sequencing problems.

\section{Appendix}
\subsection{Numerical method for the shot noise model}\label{a1}
We discuss here the numerical evaluation of the
probability of a successful read in a cluster in the
shot-noise model: $\mathds{E}[g(I)]$.
For this we need the probability density of $I$.
This random variable has a mass at 0 and a continuous
part with support on $\R^+$.
The mass at 0 is  $\mathds{P}(I = 0) = \mathds{P}(B(0,2r)) = 0) > 0$. 
For the continuous part, the interference must be conditioned
on having a point in the ball $B(0, 2r)$,
since this will ensure a non-zero interference.
Letting $P_0 = \mathds{P}(I = 0) = \mathds{P}(N(B(0,2r)) = 0)$, the expected yield is

\begin{center}
$\lambda_y(\lambda, r, l) = \lambda\mathds{E}(g(I)) = \lambda\left(P_0g(0)  + (1- P_0) \int_0^{\infty} g(x) f_{I | N(B(0,2r)) > 0}(x) dx \right)$.
\end{center}

In order to compute the integral $\int_0^{\infty} g(x) f_{I | N > 0 }(x) dx$, we need to approximate $f_{I| N > 0}$. The Fourier transform of the density can be calculated  from the Laplace transform:
\begin{align*}  
\hat{f}_{I | N > 0}(s) &= \mathds{E}(e^{-isI} | N(B(0,2r)) > 0) \\
&=  \frac{1}{1 - \mathds{P}_0} \left( \mathds{E}(e^{-isI})  - \mathds{E}(e^{-isI} | N = 0)\mathds{P}(N = 0) \right). 
\end{align*}

Then, $f_{I | N > 0}(s) = \frac{1}{1 - P_0}  \mathcal{F}^{-1}\left( \hat{f}_I(s) -  P_0 \right)$ and the mean value of interest is
\begin{center}
$\left[P_0g(0)  +  \int_0^{\infty} g(x) \mathcal{F}^{-1}\left( \hat{f}_I(s) -  P_0 \right) dx \right]$.
\end{center}
The inverse Fourier transform is approximated using the inverse fast Fourier transform (\textit{ifft} in MATLAB).
The total integral is approximated using the trapezoidal rule.



\section*{Acknowledgments}
The work of the first two authors was 
supported by a grant of the Simons Foundation
(\#197982 to UT Austin). The work of the first author was supported by the National Science Foundation Graduate Research Fellowship under Grant No. DGE-1110007.

\bibliographystyle{acm}
\bibliography{gdv}

\end{document}